\documentclass[aps,prl,twocolumn,showpacs,superscriptaddress,groupedaddress]{revtex4}  

\usepackage{graphicx}  
\usepackage{dcolumn}   
\usepackage{bm}        
\usepackage{amssymb}   
\usepackage{epstopdf}
\usepackage{epsfig,pslatex,subfigure}
\usepackage{preprintcover}  
\usepackage[hyperindex,breaklinks]{hyperref}

\PreprintCoverPaperTitle{Search for pair production of a new quark that decays to a $Z$ boson and a bottom quark with the ATLAS detector}  
\PreprintIdNumber{CERN-PH-EP-2012-073}  
\PreprintCoverAbstract{
A search is reported for the pair production of a new quark, $b'$, with at least one $b'$ decaying to a $Z$ boson and a bottom quark.  The data, corresponding to 2.0~fb$^{-1}$ of integrated luminosity, were collected from $pp$ collisions at $\sqrt{s}=7$~TeV with the ATLAS detector at the CERN Large Hadron Collider.  Using events with a \mbox{$b$-tagged} jet and a $Z$ boson reconstructed from opposite-charge electrons, the mass distribution of large transverse momentum $b'$ candidates is tested for an enhancement.  No evidence for a $b'$ signal is detected in the observed mass distribution, resulting in the exclusion at 95\% confidence level of $b'$ quarks with masses \mbox{$m_{b'}<400$~GeV} that decay entirely via \mbox{$b' \rightarrow Z+b$}.  In the case of a vector-like singlet $b'$ mixing solely with the third Standard Model generation, masses $m_{b'}<358$~GeV are excluded.}  
\PreprintJournalName{Physical Review Letters}  

\begin{document}

\title{Search for pair production of a new quark that decays to a $Z$ boson and a bottom quark \\ with the ATLAS detector}
\author{The ATLAS Collaboration}

\begin{abstract}
A search is reported for the pair production of a new quark, $b'$, with at least one $b'$ decaying to a $Z$ boson and a bottom quark.  The data, corresponding to 2.0~fb$^{-1}$ of integrated luminosity, were collected from $pp$ collisions at $\sqrt{s}=7$~TeV with the ATLAS detector at the CERN Large Hadron Collider.  Using events with a \mbox{$b$-tagged} jet and a $Z$ boson reconstructed from opposite-charge electrons, the mass distribution of large transverse momentum $b'$ candidates is tested for an enhancement.  No evidence for a $b'$ signal is detected in the observed mass distribution, resulting in the exclusion at 95\% confidence level of $b'$ quarks with masses \mbox{$m_{b'}<400$~GeV} that decay entirely via \mbox{$b' \rightarrow Z+b$}.  In the case of a vector-like singlet $b'$ mixing solely with the third Standard Model generation, masses $m_{b'}<358$~GeV are excluded.
\end{abstract}

\maketitle

The matter sector of the Standard Model (SM) consists of three generations of chiral fermions, with each generation containing a quark doublet and a lepton doublet.  A natural question is whether quarks and leptons exist beyond the third generation~\cite{FHS}.  In this Letter we present a search for the pair production of a new quark with electric charge $-1/3$, denoted $b'$, using data collected by the ATLAS experiment at the Large Hadron Collider.  New quarks appear in a variety of models that address shortcomings of the SM~\cite{FHS,Martin,BM,E6,TopPartners}.  In addition to signaling a richer matter content at high energy, their existence would impact lower-scale physics, such as altering Higgs boson ($H$) phenomenology~\cite{Higgs4Gen}, and providing new sources of CP violation potentially sufficient to generate the baryon asymmetry in the universe~\cite{BAU4Gen}. 

Several collaborations have previously searched for a chiral $b'$.  A search by D0~\cite{D0_1} for the decay $b' \rightarrow \gamma + b$ excludes $b'$ quarks with masses below $m_{Z}+m_{b}=96$~GeV.  CDF~\cite{CDF_1} searches for the decay $b' \rightarrow Z+b$ exclude masses below $m_{W} + m_{t}=256$~GeV.  These limits apply to prompt $b'$ decays.  CDF and D0 have also searched for non-prompt $b' \rightarrow Z+b$ decays~\cite{FH}, excluding, for example, $b'$ masses below 180~GeV for $c\tau = 20$~cm~\cite{D0_2}.  More recently, CDF~\cite{CDF_4}, CMS~\cite{CMS}, and ATLAS~\cite{ATLAS} have searched for the prompt charged-current decay $b' \rightarrow W+t$.  This decay mode is dominant for a chiral $b'$ with mass in excess of $m_{W} + m_{t}$, as the neutral-current modes only occur through loop diagrams~\cite{FHS}.  The ATLAS result excludes chiral $b'$ quarks with masses below 480~GeV.

Extensions to the SM often propose new quarks transforming as vector-like representations of the electroweak gauge groups~\cite{Martin,BM,E6,TopPartners}.  The decay of a vector-like $b'$ to a $Z$ boson and a bottom quark is a tree-level process with a branching ratio comparable to that of the decay $b' \rightarrow W+t$.  In particular, the branching ratios \mbox{$Wt:Zb:Hb$} approach the proportion $2:1:1$ in the limit of large $b'$ mass as a consequence of the Goldstone boson equivalence theorem~\cite{Martin,TopPartners}.  Furthermore, if a signal were observed in the $WtWt$ final state, a search for a resonant $Z+b$ signal would aid in establishing the charge of the new quark.  In light of these observations, this search explores the $Z+b$-jet final state for the presence of a $b'$ quark.

The ATLAS detector~\cite{Det} consists of particle-tracking detectors, electromagnetic and hadronic calorimeters, and a muon spectrometer.   At small radii transverse to the beamline, the inner tracking system utilizes fine-granularity pixel and microstrip detectors designed to provide precision track impact parameter and secondary vertex measurements.  These silicon-based detectors cover the pseudorapidity~\cite{DetCoord} range $|\eta|<2.5$.  A gas-filled straw tube tracker complements the silicon tracker at larger radii.  The tracking detectors are immersed in a 2~T magnetic field produced by a thin superconducting solenoid located in the same cryostat as the barrel electromagnetic (EM) calorimeter.  The EM calorimeters employ lead absorbers and utilize liquid argon as the active medium.  The barrel EM calorimeter covers $|\eta|<1.5$, and the end-cap EM calorimeters $1.4<|\eta|<3.2$.  Hadronic calorimetry in the region $|\eta|<1.7$ is achieved using steel absorbers and scintillating tiles as the active medium.  Liquid argon calorimetry with copper absorbers is employed in the hadronic end-cap calorimeters, which cover the region $1.5<|\eta|<3.2$.  

The search for the decay $b'\rightarrow Z+b$ is performed in the final state with the $Z$ boson decaying to an electron-positron pair  ($e^{+}e^{-}$) using a dataset collected in 2011 corresponding to an integrated luminosity of $1.98\pm0.07~\mathrm{fb}^{-1}$~\cite{Lumi}.  The selected events were recorded with a single-electron trigger that is over 95\% efficient for reconstructed electrons~\cite{EMID} with momentum transverse to the beam direction, $p_{\mathrm{T}}$, exceeding 25~GeV.  At least two opposite-charge electron candidates are required, each satisfying \mbox{$p_{\mathrm{T}}>25$~GeV} and reconstructed in the pseudorapidity region $|\eta|<2.47$, excluding the barrel to end-cap calorimeter transition region, \mbox{$1.37< |\eta| < 1.52$}.  In addition, the electron candidates satisfy {\it medium} quality requirements~\cite{EMID} on the reconstructed track and properties of the electromagnetic shower.  The two opposite-charge electron candidates yielding an invariant mass, $m_{ee}$, that satisfies $|m_{ee}-m_{Z}|<15$~GeV and is closest to the $Z$ boson mass define the $Z$ candidate.  Approximately 475,000 events pass the $Z \rightarrow e^{+}e^{-}$ selection criteria.

Jets are reconstructed using the anti-$k_{t}$ clustering algorithm~\cite{JetAlg} with a distance parameter of 0.4.  The inputs to the algorithm are three-dimensional clusters formed from calorimeter energy deposits.  Jets are calibrated using $p_{\mathrm{T}}$- and $\eta$-dependent factors determined from simulation and validated with data~\cite{JES}.  Jets are rejected if they do not satisfy quality criteria to suppress noise and non-collision backgrounds, as are jets whose axis is within \mbox{$\Delta R=\sqrt{(\Delta \eta)^{2}+(\Delta \phi)^{2}}=0.5$} of a reconstructed electron associated with the $Z$ candidate.  A requirement is made to ensure at least $75\%$  of the total $p_{\mathrm{T}}$ of all tracks associated with the jet be attributed to tracks also associated with the selected $pp$ collision vertex~\cite{JVF}.  Lastly, jets in this analysis are restricted to the region covered by the tracking detectors, $|\eta|<2.5$, and satisfy $p_{\mathrm{T}}>25$~GeV.  Approximately 81,000 events pass the $Z \rightarrow e^{+}e^{-}$ candidate selection and contain at least one selected jet.

The SM production of $Z$ bosons in association with jets accounts for most events passing the $Z+\geq1$~jet selection.  Two leading-order Monte Carlo (MC) generators, {\sc alpgen}~\cite{ALPGEN} and {\sc sherpa}~\cite{SHERPA}, are used to assess the background arising from this process, with {\sc alpgen} providing the baseline prediction.  A description of the generation of these samples, in particular in regard to differences between {\sc alpgen} and {\sc sherpa} in the modeling of $Z$ boson production in association with $b$-jets, is detailed in Ref.~\cite{ATLAS_Zb}.  The predictions of both are normalized such that the inclusive $Z$ boson cross section is equal to a next-to-next-to-leading-order (NNLO) calculation~\cite{FEWZ}.  All MC samples fully simulate the ATLAS detector~\cite{SIM} and are reconstructed with the same algorithms as those applied to data.  The $Z$+bottom background category comprises simulated $Z+\mathrm{jet(s)}$ events in which a generated $p_{\mathrm{T}}>5$~GeV bottom quark is matched to a selected reconstructed jet.  Similarly, events with a jet matched to a charm quark, but not a bottom quark, constitute the $Z$+charm category.  In the $Z$+light category, none of the selected jets are matched to a bottom or charm quark.  
   
Additional SM backgrounds modeled with MC events include top quark pair production ($t\bar{t}$), single top production, heavy vector boson pair (diboson) production, \mbox{$Z( \rightarrow \tau\tau)+$~jet(s)} events, and $W(\rightarrow e\nu)+$~jet(s) events.  Processes with a top quark are simulated with {\sc mc@nlo}~\cite{MCNLO1,MCNLO2}.  The $t\bar{t}$ cross section used is the {\sc hathor}~\cite{HATHOR} approximate NNLO value, while {\sc mc@nlo}~\cite{MCNLO2} values are used for the single top processes.   {\sc herwig}~\cite{HERWIG} models the contribution of diboson events, with the cross sections set by the {\sc mcfm}~\cite{MCFM1} NLO predictions.  The remaining $W/Z+$~jet(s) backgrounds are simulated with {\sc alpgen}, and normalized using single vector boson production NNLO cross sections~\cite{FEWZ}.  The multi-jet background is estimated using a data sample with both electron candidates passing {\it loose} criteria~\cite{EMID} but failing the slightly tighter {\it medium} criteria.  This sample is normalized to the difference in the inclusive $Z$ sample between the data and all other backgrounds in the region $50<m_{ee}<65$~GeV.  The small single top, diboson, $Z\rightarrow \tau\tau$, $W\rightarrow e\nu$, and multi-jet contributions are combined and denoted Other SM.  

\begin{figure}[]
\includegraphics[scale=0.42]{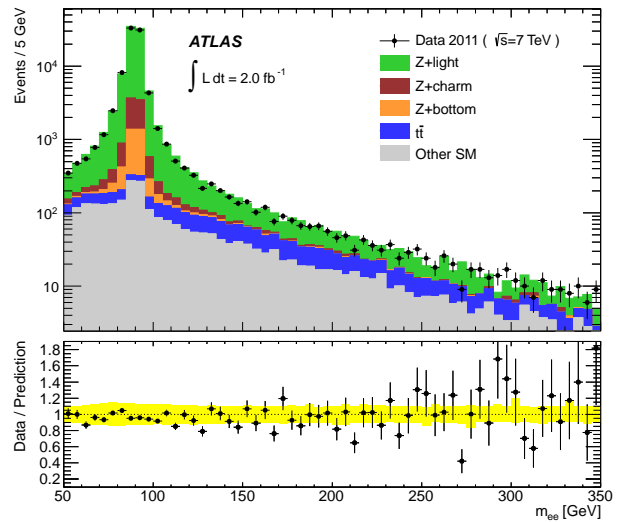}
\caption{\label{fig:1} $e^{+}e^{-}$ invariant mass distribution for events passing the $Z+\geq1$~jet selection, before imposing the $|m_{ee}-m_{Z}|<15$~GeV requirement.  The predicted contributions of the SM background sources are shown stacked.  The lower panel shows the ratio of the data to the SM prediction, and the solid yellow band denotes the systematic uncertainty on the SM prediction.}
\end{figure}

\begin{figure}[]
\includegraphics[scale=0.42]{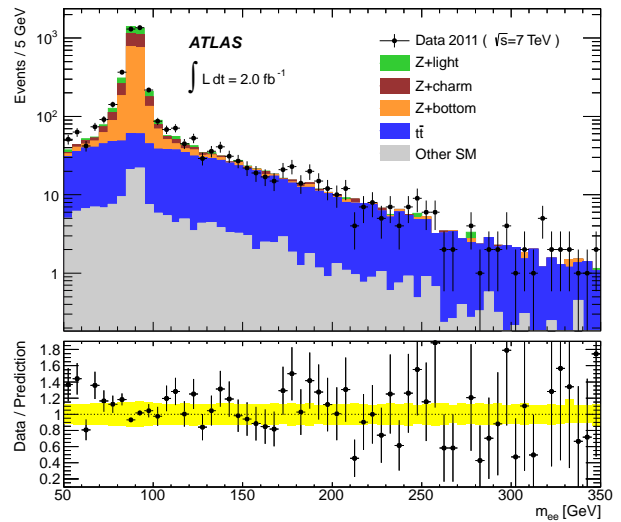}
\caption{\label{fig:2} $e^{+}e^{-}$ invariant mass distribution for events passing the $Z+\geq1$~$b$-jet selection, before imposing the $|m_{ee}-m_{Z}|<15$~GeV requirement.}
\end{figure}

\begin{table*}[t]
\begin{ruledtabular}
\begin{tabular}{c ccc}
  Source                               &    $Z+\geq1$~jet                               &  $Z+\geq1$~$b$-jet                       &   $p_{\mathrm{T}}(Zb)>150$~GeV      \\ \hline
$Z$+light                           &     $74400 \pm 7300$                     &  $\phantom{1}590 \pm 140$                  &     $19 \pm 7$       \\
$Z$+charm                        &     $5340 \pm 520$                         &  $\phantom{1}870  \pm 210$               &     $18 \pm 7$       \\
$Z$+bottom                       &     $2540 \pm 250$                         &  $1710 \pm  270$                                    &     $\phantom{1}52 \pm  17$       \\
$t\bar{t}$                               &     $\phantom{1}320 \pm 40\phantom{1}$         &  $220 \pm 40$                      &     $20 \pm  4$       \\
Other SM                              &     $1010 \pm 280$      &  $\phantom{1}70 \pm 20$                       &    $\phantom{1}1.6  \pm  0.4$     \\ \hline
Total SM                               &     $83600 \pm 8100$                     &  $3460 \pm 580$                                       &    $110 \pm 30$        \\
{\bf Data}                               &    {\bf 80519}                                  &  {\bf 3466}                                                        &   {\bf 100}      \\ \hline
$m_{b'}=350$~GeV            &     $110 \pm 12$                        &  $\phantom{1}93 \pm 11$                       &     $55 \pm 7$         \\        
$m_{b'}=450$~GeV            &     $27 \pm 3$            &             $20 \pm 2$                       &    $14 \pm 2$         \\
\end{tabular}
\end{ruledtabular}
\caption{\label{table:events} Number of predicted and observed events at three stages in the event selection.  The contributions from SM backgrounds are shown individually, as well as combined into the total SM prediction.  The uncertainties on the predicted number of events combine all sources of uncertainty.  The number of expected signal events is also listed for two representative $b'$ masses in the case where $BR(b' \rightarrow Zb)=1$.}
\end{table*}

Figure~\ref{fig:1} presents the $e^{+}e^{-}$ invariant mass distribution for events passing the $Z+\geq1$~jet selection, before imposing the $|m_{ee}-m_{Z}|<15$~GeV requirement, together with the SM prediction.  The observed and predicted number of events are listed in Table~\ref{table:events} for this and two other stages of the event selection.  Most events passing the $Z+\geq1$~jet selection arise from the $Z$+light category.  The appreciable lifetime of the $b$-hadron originating from the bottom quark in the decay \mbox{$b' \rightarrow Z+b$} provides a means to reduce this background source.  A $b$-jet tagging algorithm referred to as IP3D+SV1~\cite{Btag1} is utilized to select events with at least one $b$-jet from the $Z+\geq1$~jet sample.  The discriminant combines two likelihood variables based on the tracks associated with a jet.  The first employs the longitudinal and transverse track impact parameters, while the second utilizes properties of a reconstructed secondary vertex.  In a simulated $t\bar{t}$ sample, the requirement on the discriminant defining a $b$-jet is $60\%$ efficient for jets with a $b$-hadron, and yields a light flavor jet rejection rate of 300~\cite{Btag1}.

A total of 3,466 events satisfy the $Z+\geq1$~$b$-jet selection.  Figure~\ref{fig:2} presents the $e^{+}e^{-}$ invariant mass distribution in this sample and the SM prediction, before imposing the $|m_{ee}-m_{Z}|<15$~GeV requirement.  The accurate modeling of the mass distribution for values beyond the $Z$ boson mass supports the prediction of $t\bar{t}$ and Other SM background events.  Within the window around the $Z$ boson mass, {\sc alpgen} and {\sc sherpa} agree to within $1\%$ and $7\%$ in the prediction of the number of $Z$+light and $Z$+charm events, respectively.  However, {\sc alpgen} and {\sc sherpa} disagree in the prediction of the $Z$+bottom contribution, a fact previously reported in an ATLAS cross section measurement of $Z$ bosons produced in association with $b$-jets using a smaller dataset~\cite{ATLAS_Zb}.  The {\sc alpgen} and {\sc sherpa} $Z$+bottom predictions are scaled to account for the difference between data and all other predicted backgrounds in a subsample of the $Z+\geq1$~$b$-jet sample that contains events failing the requirement discussed below on the transverse momentum of the $b'$ candidate.  The scale factors are consistent with those measured in Ref.~\cite{ATLAS_Zb}, and the invariant mass distribution of secondary vertex tracks is used to confirm the validity of the resulting prediction for the flavor composition in the $Z+\geq1$~$b$-jet sample~\cite{ATLAS_Zb}.

Simulated $b'\bar{b}'$ events are generated for a range of $b' $ masses using {\sc madgraph}~\cite{MADGRAPH} with the G4LHC extension~\cite{Higgs4Gen}.  {\sc pythia}~\cite{pythia} performs fragmentation and hadronization of the parton-level events.  The signal cross sections are obtained with {\sc hathor}~\cite{HATHOR}, and vary from 80~pb to 30~fb over the range $m_{b'}=200-700$~GeV.  In each sample, one $b'$ decays in the mode $b' \rightarrow Z+b$, with the $Z$ boson decaying via $Z\rightarrow e^{+}e^{-}$.  Two separate samples are produced for each mass value, with the other $b'$ decaying either via $b' \rightarrow Z+b$ or $b' \rightarrow W+t$, and with all decay modes of the $Z$ and $W$ bosons allowed.  The factor \mbox{$\beta = 2\times BR(b'\rightarrow Zb)-BR(b'\rightarrow Zb)^{2}$} characterizes the fraction of signal events with at least one $b'\rightarrow Z+b$ decay as a function of the branching ratio.  The case $\beta=1$ is equivalent to previous measurements~\cite{CDF_1} which assumed $BR(b'\rightarrow Zb)=1$.  The case of a vector-like singlet (VLS) mixing solely with the third SM generation is also considered by computing $\beta$ as a function of $b'$ mass~\cite{TopPartners}.   Over the range $m_{b'}=200-700$~GeV, $\beta$ varies from $0.9$ to $0.5$.  A SM~Higgs of mass $125$~GeV is assumed.

\begin{figure}[t]
\includegraphics[scale=0.42]{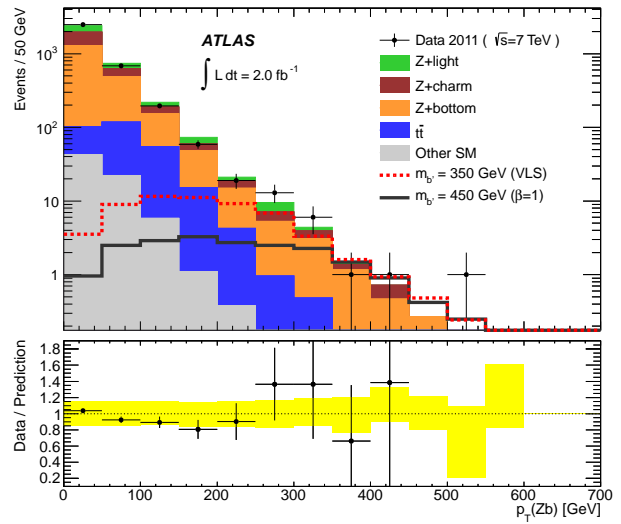}
\caption{\label{fig:3} Transverse momentum distribution of the $b'$ candidate in events passing the $Z+\geq1$~$b$-jet selection.  The predicted contributions of the SM background sources are stacked, while the distributions for the two signal scenarios described in the text are overlaid.}
\end{figure}

The $b'$ candidate is formed from the  $e^{+}e^{-}$ pair and the highest $p_{\mathrm{T}}$ $b$-jet.  The mass of the $b'$ candidate, $m(Zb)$, is the discriminant distinguishing the background-only and signal-plus-background hypotheses.  In $b'$ pair production, the new quarks are typically produced with large transverse momentum, $p_{\mathrm{T}}(Zb)$.  Therefore, a $p_{\mathrm{T}}(Zb)>150$~GeV requirement is applied to increase the signal sensitivity.  Figure~\ref{fig:3} presents the $p_{\mathrm{T}}(Zb)$ distribution for data and the predicted SM backgrounds.  Additionally, the signal distribution is overlaid for a $b'$ mass of 350~GeV, assuming the VLS scenario value $\beta=0.63$, and for a mass of 450~GeV, assuming $\beta=1$.

\begin{figure}[t]
\includegraphics[scale=0.42]{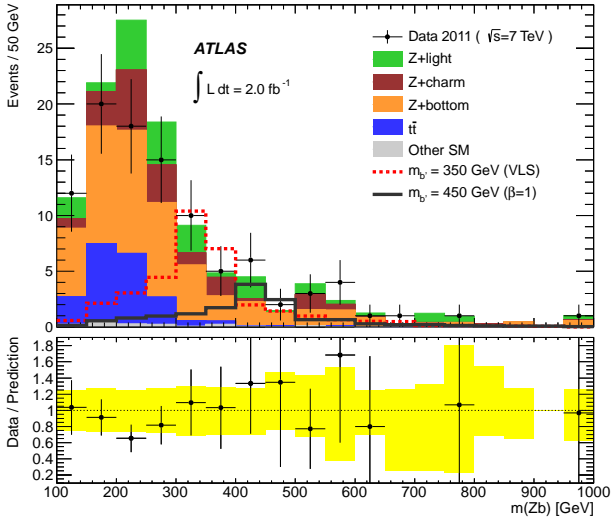}
\caption{\label{fig:4} Mass distribution of the $b'$ candidate in events passing the $Z+\geq1$~$b$-jet selection and satisfying $p_{\mathrm{T}}(Zb)>150$~GeV.  The highest mass bin also includes the data and prediction for  $m(Zb)>1$~TeV.}
\end{figure}

The fraction of signal events passing all requirements varies from $7\%$ to $43\%$ between $m_{b'}=200-700$~GeV,  assuming \mbox{$\beta=1$}, with the efficiency to pass the minimum $p_{\mathrm{T}}(Zb)$ requirement contributing most to the degree of variation.  The requirement \mbox{$p_{\mathrm{T}}(Zb)>150$~GeV} was determined by assessing the signal sensitivity for different minimum $p_{\mathrm{T}}(Zb)$ values,  as quantified by the expected cross section exclusion limit.  The limit is computed using a binned Poisson likelihood ratio test~\cite{Collie} of the $m(Zb)$ distribution for different $m_{b'}$ hypotheses.  Pseudo-experiments are generated according to the background-only and signal-plus-background hypotheses, and incorporate the impact of systematic uncertainties.  The cross section limit is evaluated using the $\mathrm{CL}_{s}$ modified frequentist approach~\cite{Collie}.

The impact of each systematic uncertainty on the normalization and shape of the $m(Zb)$ distribution is assessed for each SM background source and the expected $b' $ signal.  The fractional  uncertainty on the total number of background events passing the $p_{\mathrm{T}}(Zb)>150$ GeV requirement is 27\%.  Significant contributions arise from uncertainties in the $p_{\mathrm{T}}(Zb)$ distribution shape in $Z+\mathrm{jet(s)}$ events.  Such sources of uncertainty include the renormalization and factorization scale choice (14\%, evaluated using {\sc mcfm}~\cite{MCFM2}), shape differences observed between {\sc alpgen} and {\sc sherpa} (12\%), and variations in the degree of initial and final state QCD radiation (9\%).  The uncertainty in the efficiency of the $b$-tagging requirement contributes an additional 12\%.  Other sources of uncertainty contributing at the level of 6\% or less include the jet energy scale~\cite{JES} , parton distribution functions (PDF), MC sample sizes, electron identification efficiency, $Z$ boson cross section, luminosity, $b$-jet mis-tag rate, $t\bar{t}$ cross section, jet energy resolution, trigger efficiency, and the Other SM event yield.  Most of the above uncertainties, with the notable exception of the $p_{\mathrm{T}}(Zb)$ modeling uncertainties in $Z+\mathrm{jet(s)}$ events, contribute to the total uncertainty on the signal normalization, which varies between 11\% and 14\% depending on the $b'$ mass.

\begin{figure}[]
\includegraphics[scale=0.43]{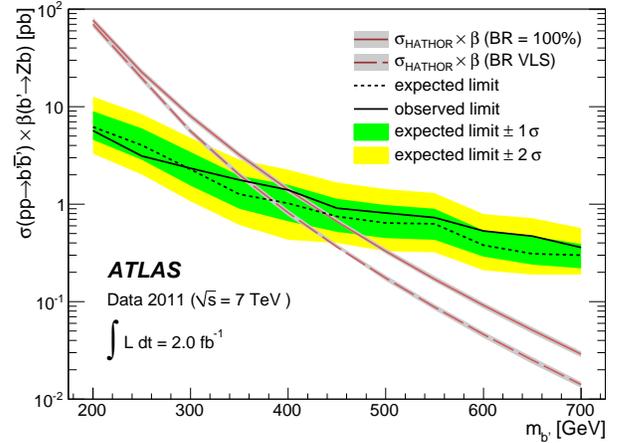}
\caption{\label{fig:5} The expected and observed $95\%$ C.L. cross section limits as a function of $b'$ mass.  The signal cross section is shown with uncertainties arising from PDFs and renormalization and factorization scale choice. The prediction is also multiplied by the $\beta$ factors described in the text.}
\end{figure}

Figure~\ref{fig:4} presents the $b'$ candidate mass distribution after requiring $p_{\mathrm{T}}(Zb)>150$~GeV and the predicted SM background.  The distributions for the signal scenarios depicted in Fig.~\ref{fig:3} are shown overlaid.  The data are in agreement with the SM prediction over the full range of $m(Zb)$ values.  In the absence of evidence of an enhancement, 95\% confidence level (C.L.) cross section exclusion limits are derived.  Figure~\ref{fig:5} presents the expected and observed cross section limits as a function of $m_{b'}$, computed under the assumption $\beta =1$.  The expected cross section limit was checked to be stable to within $15\%$ over the full mass range considered using the signal samples in which one $b'$ quark decays via $b' \rightarrow Z+b$ and the other decays via $b' \rightarrow W+t$.  The approximate NNLO $b'\bar{b}'$ cross section prediction is shown multiplied by $\beta=1$, as well as by the VLS $\beta$ value, with the shaded region representing the total uncertainty arising from PDF uncertainties and the  factorization and renormalization scale choice.  From the intersection of the observed cross section limit and the theoretical prediction, $b'$ quarks with masses \mbox{$m_{b'}<400$~GeV} decaying entirely via \mbox{$b' \rightarrow Z+b$} are excluded at 95\% C.L., representing a significant improvement with respect to the previous best limit of 268~GeV~\cite{CDF_1}.  In the case of a vector-like singlet $b'$ mixing solely with the third SM generation, masses $m_{b'}<358$~GeV are excluded.

In conclusion, a search with 2.0~fb$^{-1}$ of ATLAS data is presented for $b'$ quark pair production, with at least one $b'$ decaying to a $Z$ boson and a bottom quark.  This decay mode is particularly relevant in the context of vector-like quarks and is an essential complement to searches in the mode with both $b'$ decaying to a $W$ boson and a top quark.  No evidence for a $b'$ is observed in the $Z+b$-jet final state, and new limits are derived on the mass of a $b'$ quark decaying via $b' \rightarrow Z+b$.

We thank CERN for the very successful operation of the LHC, as well as the support staff from our institutions without whom ATLAS could not be operated efficiently.

We acknowledge the support of ANPCyT, Argentina; YerPhI, Armenia; ARC, Australia; BMWF, Austria; ANAS, Azerbaijan; SSTC, Belarus; CNPq and FAPESP, Brazil; NSERC, NRC and CFI, Canada; CERN; CONICYT, Chile; CAS, MOST and NSFC, China; COLCIENCIAS, Colombia; MSMT CR, MPO CR and VSC CR, Czech Republic; DNRF, DNSRC and Lundbeck Foundation, Denmark; EPLANET and ERC, European Union; IN2P3-CNRS, CEA-DSM/IRFU, France; GNAS, Georgia; BMBF, DFG, HGF, MPG and AvH Foundation, Germany; GSRT, Greece; ISF, MINERVA, GIF, DIP and Benoziyo Center, Israel; INFN, Italy; MEXT and JSPS, Japan; CNRST, Morocco; FOM and NWO, Netherlands; RCN, Norway; MNiSW, Poland; GRICES and FCT, Portugal; MERYS (MECTS), Romania; MES of Russia and ROSATOM, Russian Federation; JINR; MSTD, Serbia; MSSR, Slovakia; ARRS and MVZT, Slovenia; DST/NRF, South Africa; MICINN, Spain; SRC and Wallenberg Foundation, Sweden; SER, SNSF and Cantons of Bern and Geneva, Switzerland; NSC, Taiwan; TAEK, Turkey; STFC, the Royal Society and Leverhulme Trust, United Kingdom; DOE and NSF, United States of America.

The crucial computing support from all WLCG partners is acknowledged gratefully, in particular from CERN and the ATLAS Tier-1 facilities at TRIUMF (Canada), NDGF (Denmark, Norway, Sweden), CC-IN2P3 (France), KIT/GridKA (Germany), INFN-CNAF (Italy), NL-T1 (Netherlands), PIC (Spain), ASGC (Taiwan), RAL (UK) and BNL (USA) and in the Tier-2 facilities worldwide.

\onecolumngrid
\clearpage
\begin{flushleft}
{\Large The ATLAS Collaboration}

\bigskip

G.~Aad$^{\rm 48}$,
B.~Abbott$^{\rm 112}$,
J.~Abdallah$^{\rm 11}$,
S.~Abdel~Khalek$^{\rm 116}$,
A.A.~Abdelalim$^{\rm 49}$,
A.~Abdesselam$^{\rm 119}$,
O.~Abdinov$^{\rm 10}$,
B.~Abi$^{\rm 113}$,
M.~Abolins$^{\rm 89}$,
O.S.~AbouZeid$^{\rm 159}$
% ...
\bigskip

$^{1}$ University at Albany, Albany NY, United States of America\\
% ...
$^{10}$ Institute of Physics, Azerbaijan Academy of Sciences, Baku, Azerbaijan\\
$^{11}$ Institut de F\'isica d'Altes Energies and Departament de F\'isica de la Universitat Aut\`onoma  de Barcelona and ICREA, Barcelona, Spain\\
% ...
$^{48}$ Fakult\"{a}t f\"{u}r Mathematik und Physik, Albert-Ludwigs-Universit\"{a}t, Freiburg i.Br., Germany\\
$^{49}$ Section de Physique, Universit\'e de Gen\`eve, Geneva, Switzerland\\
% ...
$^{89}$ Department of Physics and Astronomy, Michigan State University, East Lansing MI, United States of America\\
% ...
$^{112}$ Homer L. Dodge Department of Physics and Astronomy, University of Oklahoma, Norman OK, United States of America\\
$^{113}$ Department of Physics, Oklahoma State University, Stillwater OK, United States of America\\
% ...
$^{116}$ LAL, Univ. Paris-Sud and CNRS/IN2P3, Orsay, France\\
% ...
$^{119}$ Department of Physics, Oxford University, Oxford, United Kingdom\\
% ...
$^{159}$ Department of Physics, University of Toronto, Toronto ON, Canada\\
% ...
\end{flushleft}


\begin{thebibliography}{99}
\bibitem{DetCoord} ATLAS uses a right-handed coordinate system with its origin at the nominal interaction point (IP) in the center of the detector and the $z$-axis coinciding with the axis of the beam pipe. The $x$-axis points from the IP to the center of the LHC ring, and the $y$-axis points upward.  Cylindrical coordinates $(r,\phi)$ used in the transverse plane, $\phi$ being the azimuthal angle around the beam pipe.  The pseudorapidity is defined in terms of the polar angle, $\theta$ as $\eta = -\ln \tan( \theta / 2)$.
\end{thebibliography}
\end{document}